\documentclass[12pt,reqno]{amsart}
\usepackage{amssymb,latexsym,amsmath}
\usepackage{amscd,amsfonts}
\usepackage{epsfig}

\newtheorem{theorem}{Theorem}

\newtheorem{proposition}{Proposition}
\newtheorem{corollary}{Corollary}

\theoremstyle{definition}
\newtheorem{remark}{Remark}

\numberwithin{equation}{section}

\begin{document}
\title{Thermodynamic length in a 2-D thermodynamic state space}
\author{Manuel Santoro}\address{Department of Mathematics and
Statistics, Portland State University, PO Box 751, Portland, OR
97207-0751, USA} \email{emanus@pdx.edu}
\begin{abstract}
The main goal of this paper is to reach an explicit formulation
and possible interpretation of thermodynamic length in a
thermodynamic state space with two degrees of freedom. Using the
energy and entropy metric in a general form, we get explicit
results about thermodynamic length along isotherms, its relation
with work and with speed of sound. We also look at the relation
between the determinants of the energy and entropy metrics and
find that they differ by a factor of $T^{4}$.
\end{abstract}\bigskip
\maketitle
\section{Introduction}
\par
The concept of thermodynamic length defined by using either the
energy or the entropy metric were introduced by F.Weinhold,$^{10}$
and G.Ruppeiner,$^{4}$ and later used by P. Salamon,$^{5-8}$, R.S.
Berry,$^{5,7,8}$, J. Nulton,$^{6}$, E. Ihrig,$^{6}$ and others to
study thermodynamic length. In particular, Salamon, Nulton and
Ihrig $^{6}$ found that thermodynamic length using the entropy and
energy metrics differs by a factor of the square root of mean
temperature during a thermodynamic process and that,
dimensionally, length is the square root of energy or it has the
dimension of velocity if internal energy is given per unit
mass.$^{7}$ It is also shown explicitly an expression for the
length of an ideal gas at constant volume, pressure and entropy.
We go a step further showing explicitly a possible interpretation
of length in a general two dimensional thermodynamic state space
with both metrics along isotherms. We find an expression for
thermodynamic length in terms of extensive variables both in
integral and differential form, its relation with the determinant
of the metrics and the speed of sound. We also show the case when
length is a measure of work done by the system and an interesting
relation between the determinants of the two metrics.
\par
\section{Thermodynamic length with the energy metric}
\par
The thermodynamic length between two points $a_{0}$ and $a_{1}$ in
thermodynamic state space is given by the following equation
\par
\begin{equation}
L_{a_{0}a_{1}}=\int^{a_{1}}_{a_{0}}{[\sum_{i,j}\eta_{ij}dX_{i}dX_{j}]^{\frac{1}{2}}}
\end{equation}
\par
where $\eta_{ij}$ are elements of the thermodynamic metric and
$X_{i}$ represent independent coordinates in thermodynamic state
space. We are going to study thermodynamic length with both the
energy and entropy metric in a two dimensional state space with
the purpose of relating length to the "degeneracy" of the
thermodynamical state space and, for a restricted class of
systems, to the concept of work.\par This manuscript will unfold
in a sequence of three main points. The first is that, as we shall
see, thermodynamic length of a thermodynamical system with two
degrees of freedom goes to zero as the system reaches its critical
point and, therefore, as curvature blows up. The second is that
also the speed of sound goes to zero close to criticality and the
third is that for Ideal and "quasi-Ideal" thermodynamical
systems,(see Remark 1 below), thermodynamic length is proportional
to work as long as we keep the temperature of the (reversible)
process constant. In particular, it would be reasonable to think
that length could quantify the amount of work done by a
"quasi-Ideal" system to keep its equilibrium.
\par
\mathstrut
 We first consider the energy representation in which
the molar internal energy is given by $u=u(s,v)$ where $s$, the
molar entropy, and $v$, the molar volume, are the two independent
variables.
\par
Then the metric for such a system is given by,$^{3}$
\par
\begin{equation}
\eta_{ij_{u}}=\frac{1}{c_{v}}
\begin{pmatrix}
 T & -\frac{T\alpha}{k_{T}}\\
 -\frac{T\alpha}{k_{T}} & \frac{c_{p}}{vk_{T}}
\end{pmatrix}
\end{equation}
\par
where
\par
\begin{enumerate}
\item $c_{v}$ is the molar heat capacity at constant volume:
\begin{equation}
c_{v}=T(\frac{\partial s}{\partial T})_{v}\qquad,
\end{equation}
\item $c_{p}$ is the molar heat capacity at constant pressure:
\begin{equation}
c_{p}=T(\frac{\partial s}{\partial T})_{p}\qquad,
\end{equation}
\item $\alpha$ is the thermal coefficient of expansion:
\begin{equation}
\alpha=\frac{1}{v}(\frac{\partial v}{\partial T})_{p}\qquad,
\end{equation}
\item $\kappa_{T}$ is the isothermal compressibility:
\begin{equation}
\kappa_{T}=-\frac{1}{v}(\frac{\partial v}{\partial p})_{T}\qquad.
\end{equation}
\end{enumerate}
\par
Therefore the thermodynamic length in the energy metric with
entropy and volume becomes
\par
\[
L=\int{[\frac{T}{c_{v}}(ds)^{2}-2\frac{T\alpha}{\kappa_{T}c_{v}}dsdv+\frac{c_{p}}{v\kappa_{T}c_{v}}(dv)^{2}]^{\frac{1}{2}}}
\]
\par
\begin{equation}
=\int\sqrt{{\frac{T}{c_{v}}[ds-\frac{\alpha}{\kappa_{T}}dv]^{2}+\frac{1}{v\kappa_{T}}(dv)^{2}}}
\end{equation}
\par
In general, molar entropy and volume could be given parametrically
as $s=s(\xi)$, $v=v(\xi)$. Then the thermodynamic length would be
given by,$^{7}$
\par
\begin{equation}
L=\int^{\xi_{f}}_{\xi_{i}}[\frac{T}{c_{v}}(\frac{ds}{d\xi})^{2}-2\frac{T\alpha}{\kappa_{T}c_{v}}\frac{ds}{d\xi}\frac{dv}{d\xi}+\frac{c_{p}}{v\kappa_{T}c_{v}}(\frac{dv}{d\xi})^{2}]^\frac{1}{2}d\xi
\end{equation}
\par
Generalizing the work done by Salamon, Andresen, Gait and
Berry$^{7}$, the length for a reversible process at constant molar
entropy reduces to
\par
\begin{equation}
L^{s}=\int{\sqrt{\frac{c_{p}}{c_{v}v\kappa_{T}}}dv}=\int{\sqrt{\eta_{22}}dv}\qquad.
\end{equation}
\par
but, since
$\det{(\eta_{ij})_{u}}=\frac{T}{\kappa_{T}vc_{v}}=-\frac{T}{c_{v}}(\frac{\partial{p}}{\partial{v}})_{T}$,
then $\kappa_{T}=\frac{T}{vc_{v}\det{(\eta_{ij})_{u}}}$ and we
have
\par
\begin{equation}
L^{s}=\int{\sqrt{{\frac{c_{p}}{T}\det({\eta_{ij}})_{u}}}dv}
\end{equation}
\par
We can also look at the integration $(2.7)$ with respect to molar
volume being constant. In this case we get,
\par
\begin{equation}
L^{v}=\int{\sqrt{\frac{T}{c_{v}}}ds}=\int{\sqrt{\eta_{11}}ds}
\end{equation}
\par
Like $(2.10)$, we also have
\par
\begin{equation}
L^{v}=\int{\sqrt{v\kappa_{T}\det({\eta_{ij}})_{u}}ds}
\end{equation}
\par
It is reasonable to think that the determinant of the matrix would
be a "measure" of how close a system is to the critical point.
Indeed, if $\det{(\eta_{ij})_{u}}=0$ then we have a curve of
degeneracy along which curvature blows up and , therefore,
thermodynamical phase space become extremely curved. So, whenever
this event happens, length approaches zero and, as below is shown,
the speed of sound goes to zero.
\par
We could also look at length at constant pressure. Then we get
\par
\begin{equation}
L^{p}=\int{\sqrt{\frac{c_{p}}{Tv^{2}\alpha^{2}}}dv}
\end{equation}
\par
\mathstrut\ Let's look, now, at the relation between length and
speed of sound. Let's denote with $\nu^{i}_{sound}$ and
$\nu^{a}_{sound}$ respectively the speed of sound in terms of
isothermal and adiabatic compressibility $\kappa_{T}$ and
$\kappa_{S}$. We can, now, look at the relation between speed of
sound and the determinant of the matrix( and therefore curvature)
being aware of the fact that $\kappa_{S}$ is used the most. So,
since $\nu^{i}_{sound}=\sqrt{\frac{1}{\kappa_{T}\rho}}$, we have
the following
\par
\begin{equation}
\nu^{i}_{sound}=\sqrt{\frac{vc_{v}\det(\eta_{ij})_{u}}{T\rho}}
\end{equation}
\par
and, since $\nu^{a}_{sound}=\sqrt{\frac{1}{\kappa_{S}\rho}}$, with
$\kappa_{S}=\frac{c_{v}}{c_{p}}\kappa_{T}$, then
\par
\begin{equation}
\nu^{a}_{sound}=\sqrt{\frac{vc_{p}\det(\eta_{ij})_{u}}{T\rho}}
\end{equation}
\par
with $\rho$ being the density of the gas. It easily follows then
\par
\begin{equation}
\nu^{i}_{sound}=(\sqrt{\frac{c_{p}}{c_{v}}})\nu^{a}_{sound}
\end{equation}
\par
We can write the thermodynamic length in terms of speed of sound,
namely,
\par
\begin{equation}
L=\int{\nu^{i}_{sound}\sqrt{\frac{c_{p}\rho}{c_{v}v}}dv}=\int{\nu^{a}_{sound}\sqrt{\frac{\rho}{v}}dv}=\int{\nu^{i}_{sound}\sqrt{\frac{T\kappa_{T}\rho}{c_{v}}}ds}=\int{\nu^{a}_{sound}\sqrt{\frac{T\kappa_{T}\rho}{c_{p}}}ds}
\end{equation}
\par
Therefore, we get
\par
\begin{equation}
\frac{dL}{dv}=\nu^{i}_{sound}\sqrt{\frac{c_{p}\rho}{c_{v}v}}=\nu^{a}_{sound}\sqrt{\frac{{\rho}}{v}}
\end{equation}
\par
and
\par
\begin{equation}
\frac{dL}{ds}=\nu^{i}_{sound}\sqrt{\frac{T\kappa_{T}\rho}{c_{v}}}=\nu^{a}_{sound}\sqrt{\frac{T\kappa_{T}\rho}{c_{p}}}
\end{equation}
\par
It is evident that
\par
\[
\frac{dL}{dv}\geq{0}\qquad\frac{dL}{ds}\geq{0}
\]
\par
Moreover, since
$\frac{c_{p}}{c_{v}v\kappa_{T}}=(\frac{\partial^{2}{u}}{\partial{v^{2}}})_{s}=\eta_{22}$
and
$\frac{T}{c_{v}}=(\frac{\partial^{2}{u}}{\partial{s^{2}}})_{v}=\eta_{11}$,
then $(2.9)$ and $(2.11)$ tell us that
\par
\begin{equation}
(\frac{dL}{dv})^{2}=(\frac{\partial^{2}{}u}{\partial{v^{2}}})_{s}=\eta_{22}
\end{equation}
\par
and
\par
\begin{equation}
(\frac{dL}{ds})^{2}=(\frac{\partial^{2}{}u}{\partial{s^{2}}})_{v}=\eta_{11}
\end{equation}
\par
\begin{proposition}
The molar internal energy of the system has the form
$u=a(s)v+b(s)$ if and only if the thermodynamic length L is
independent on volume density;($v$ is the isotropic direction).
\end{proposition}
\par
\begin{proposition}
The molar internal energy of the system has the form
$u=c(v)s+d(v)$ if and only if the thermodynamic length L is
independent on molar entropy; ($s$ is the isotropic direction).
\end{proposition}
\par
Proof. Immediate from $(2.20)$ and $(2.21)_{\blacksquare}$.
\par
\mathstrut
 It is interesting to look now at the relation between
thermodynamic length and work and we'll try to give a possible
interpretation of what length would mean in a thermodynamical
system along isotherms . It is shown, indeed, by the following
theorem and remark that if our system is ideal or "quasi-ideal"
then thermodynamic length is work just along isotherms.
\par
\begin{theorem}
Let's consider the Ideal gas whose pressure is given by
\par
\begin{equation}
|(\frac{\partial{u}}{\partial{v}})_{s}|=|p|=\frac{RT}{v}
\end{equation}
\par
then, along isotherms,
\par
\begin{equation}
L^{s}=\frac{1}{\sqrt{{RT}}}W
\end{equation}
\par
where W is work.
\end{theorem}
\par
Proof.
\par
Consider the case in which
\par
\begin{equation}
(\frac{\partial^{2}{u}}{\partial{v}^{2}})_{s}=\frac{1}{RT}[(\frac{\partial{u}}{\partial{v}})_{s}]^{2}
\end{equation}
\par
where u is the molar internal energy. Naturally, $(2.24)$ is
equivalent to
\par
\begin{equation}
(\frac{\partial{p}}{\partial{v}})_{s}+\frac{1}{RT}p^{2}=0
\end{equation}
\par
since
$(\frac{\partial^{2}{u}}{\partial{v}^{2}})_{s}=-(\frac{\partial{p}}{\partial{v}})_{s}$
and $(\frac{\partial{u}}{\partial{v}})_{s}=-p$. Now, $(2.25)$ is a
separable first order differential equation whose solution is
given by
\par
\[
p=\frac{RT}{v}
\]
\par
As solution of $(2.24)$ we have $u=-RT\ln{|v|}$. Then, by $(2.9)$
\par
\begin{equation}
L^{s}=\int{\sqrt{({\frac{\partial^{2}{u}}{\partial{v}^{2}}})_{s}}}dv=\int{\sqrt{{\frac{1}{RT}(\frac{\partial{u}}{\partial{v}})^{2}_{s}}}}dv=\int{\sqrt{{\frac{1}{RT}}}|(\frac{\partial{u}}{\partial{v}})_{s}|}dv=\sqrt{{\frac{1}{RT}}}\int|{p}|dv=\sqrt{{\frac{1}{RT}}}W
_{\blacksquare}
\end{equation}
\par
\begin{remark}
Note that the statement of Theorem $1$ is true also for a more
general case in which we consider the volume occupied by molecules
inside the system. Namely the "quasi-Ideal" case in which pressure
is expressed in the form $p=\frac{RT}{v-b}$.
\end{remark}
\par
\begin{remark}
It is also extremely important to note that, while
$\eta_{22}=\frac{c_{p}p}{c_{v}v}$ for a general two dimensional
Ideal Gas, it becomes $\eta_{22}=\frac{RT}{v^{2}}$ along
isotherms.
\end{remark}
\par
So, this theorem gives us a first understanding of the relation
between thermodynamic length and work in explicit form along and
just along isotherms.
\par
\begin{corollary}
Thermodynamic length measures the amount of work done by a
"quasi-ideal" system along isotherms.
\end{corollary}
\par
By quasi-ideal we mean that $b$ might as well be different than
zero. Therefore the system is not completely ideal.
\par
\begin{remark}
It is interesting to note that equation $(2.24)$ is of the form
$u_{vv}=\frac{1}{RT}(u_{v})^{2}$ with u parametrized by molar
volume. As the volume decreases, the system applies a resistance
in the opposite (positive) direction by $\frac{1}{RT}$ (constant).
The length, then, would probably measure the amount of work done
by the system to keep its equilibrium state, always along
isotherms.
\end{remark}
\section{Thermodynamic length with the entropy metric}
\par
Let's now discuss what happens when we look at the entropy metric.
Consider $s=s(u,v)$ where u is the energy of the system per unit
mole. In this case molar internal energy and molar volume are the
independent variables. If we let
$\eta_{ij_{s}}=-\frac{\partial^{2}s(X)}{\partial{X_{i}}{\partial{X_{j}}}}$,
then the metric is given by,(see also the paper by Mrugala$^{3}$),
\par
\begin{equation}
\eta_{ij_{s}} =\frac{1}{c_{v}T^{2}}
\begin{pmatrix}
1 & -(\frac{T\alpha-\kappa_{T}p}{\kappa_{T}}) \\
-(\frac{T\alpha-\kappa_{T}p}{\kappa_{T}}) &
(\frac{T\alpha-\kappa_{T}p}{\kappa_{T}})^{2}+\frac{c_{v}T}{v\kappa_{T}}
\end{pmatrix}
\end{equation}
\par
Then the thermodynamical length is given by
\par
\[
L=\int{[\frac{1}{c_{v}T^{2}}(du)^{2}-\frac{2}{T^{2}}(\frac{T\alpha-\kappa_{T}p}{\kappa_{T}c_{v}})dudv+[\frac{1}{c_{v}}(\frac{T\alpha-\kappa_{T}p}{T\kappa_{T}})^{2}+\frac{1}{v\kappa_{T}T}(dv)^{2}]]^{\frac{1}{2}}}
\]
\par
\begin{equation}
=\int{[\frac{1}{c_{v}T^{2}}[du-(\frac{T\alpha-\kappa_{T}p}{\kappa_{T}})dv]^{2}+\frac{1}{v\kappa_{T}T}(dv)^{2}]^{\frac{1}{2}}}
\end{equation}
\par
Then, thermodynamic length at constant molar internal energy and
constant molar volume is given by
\par
\begin{equation}
L^{u}=\int{\sqrt{\frac{1}{c_{v}T^{2}}(\frac{T\alpha-\kappa_{T}p}{T\kappa_{T}})^{2}+\frac{1}{v\kappa_{T}T}}dv}=\int{\sqrt{\eta_{22}}dv}
\end{equation}
\par
and
\par
\begin{equation}
L^{v}=\int{\sqrt{\frac{1}{c_{v}T^{2}}}du}=\int{\sqrt{\eta_{11}}du}
\end{equation}
\par
Expressing the thermodynamic length in terms of the determinant of
the entropy metric which is given by
$\det(\eta_{ij})_{s}=\frac{1}{vT^{3}c_{v}\kappa_{T}}=-\frac{1}{c_{v}T^{3}}(\frac{\partial{p}}{\partial{v}})_{T}$,
we get
\par
\begin{equation}
L^{u}=\int{\sqrt{[\frac{v}{\kappa_{T}T}(T\alpha-p\kappa_{T})^{2}+T^{2}c_{v}]\det({\eta_{ij}})_{s}}dv}
\end{equation}
\par
and
\par
\begin{equation}
L^{v}=\int{\sqrt{Tv\kappa_{T}\det({\eta_{ij}})_{s}}du}
\end{equation}
\par
Moreover, we can consider the speed of sound both isothermal and
adiabatic with respect to the entropy metric
\par
\begin{equation}
\nu^{i}_{sound}=\sqrt{\frac{vc_{v}T^{3}\det(\eta_{ij})_{s}}{\rho}}
\end{equation}
\par
and
\par
\begin{equation}
\nu^{a}_{sound}=\sqrt{\frac{vc_{p}T^{3}\det(\eta_{ij})_{s}}{\rho}}
\end{equation}
\par
and write the thermodynamic length in terms of speed of sound
\par
\begin{equation}
L=\int{\nu^{i}_{sound}\sqrt{\frac{\rho}{Tv}[1+\frac{v}{T^{3}c_{v}\kappa_{T}}(T\alpha-p\kappa_{T})^{2}]}dv}=\int{\nu^{i}_{sound}\sqrt{\frac{\rho\kappa_{T}}{T^{2}c_{v}}}du}
\end{equation}
\par
Naturally $\frac{dL}{dv}\geq{0}$ and $\frac{dL}{du}\geq{0}$.
\par
As for the energy metric,( see $(2.20)$ and $(2.21)$), we have the
following propositions:
\par
\begin{proposition}
The molar entropy of the system has the form $s=a(u)v+b(u)$ if and
only if the thermodynamic length L is independent on volume
density; ($v$ is the isotropic direction).
\end{proposition}
\par
\begin{proposition}
The molar entropy of the system has the form $s=c(v)u+b(v)$ if and
only if the thermodynamic length L is independent on molar
internal energy; ($u$ is the isotropic direction).
\end{proposition}
\par
Also with the entropy metric we come to the conclusion that
thermodynamic length is somehow related to work just along
isotherms in case of Ideal and quasi-Ideal Gas. The following
theorem shows the Ideal case while the quasi-ideal is trivial as
in the case of the energy metric.
\par
\begin{theorem}
Let's consider the Ideal Gas with
\par
\begin{equation}
|(\frac{\partial{s}}{\partial{v}})_{u}|=|(\frac{p}{T})|=\frac{R}{v}
\end{equation}
\par
then, along isotherms,
\par
\begin{equation}
L^{u}=\sqrt{\frac{1}{RT^{2}}}W
\end{equation}
\par
where W is work.
\end{theorem}
\par
Proof.
\par
The following separable first order differential equation
$(\frac{\partial({p/T})}{\partial{v}})_{u}+\frac{1}{R}(\frac{p}{T})^{2}=0$
which is equivalent to
\par
\begin{equation}
(\frac{\partial^{2}{s}}{\partial{v}^{2}})_{u}=-\frac{1}{R}[(\frac{\partial{s}}{\partial{v}})_{u}]^{2}
\end{equation}
\par
has solution $p=\frac{RT}{v}$.
\par
Therefore $(3.3)$ becomes
\par
\begin{equation}
L^{u}=\int{\sqrt{({\frac{\partial^{2}{s}}{\partial{v}^{2}}})_{u}}}dv=\int{\sqrt{{\frac{1}{R}(\frac{\partial{s}}{\partial{v}})^{2}_{u}}}}dv=\int{\sqrt{{\frac{1}{R}}}|(\frac{\partial{s}}{\partial{v}})_{u}|}dv=\sqrt{{\frac{1}{R}}}\int{|\frac{p}{T}|}dv=\sqrt{{\frac{1}{RT^{2}}}}W
_{\blacksquare}
\end{equation}
\par
\begin{remark}
Note that also in this case, the statement of Theorem $2$ is true
for a more general case in which the volume occupied by molecules
inside the system is considered. Namely the "quasi-Ideal" case.
\end{remark}
\par
 \mathstrut
\par
\section{Relation between $\det(\eta_{ij})_{u}$ and
$\det(\eta_{ij})_{s}$}
\par
We have found that the determinants of the two matrices $(2.2)$
and $(3.1)$ were given by
\par
\begin{equation}
\det{(\eta_{ij})_{u}}=\frac{T}{\kappa_{T}vc_{v}}=-\frac{T}{c_{v}}(\frac{\partial{p}}{\partial{v}})_{T}
\end{equation}
and
\par
\begin{equation}
\det{(\eta_{ij})_{s}}=\frac{1}{vT^{3}c_{v}\kappa_{T}}=-\frac{1}{c_{v}T^{3}}(\frac{\partial{p}}{\partial{v}})_{T}
\end{equation}
\par
Note that in case the entropy metric is considered to be the
positive second partial derivative of entropy density with respect
to the extensive variables than the signs of such expressions
change. Now, comparing the two determinants we get the interesting
result
\par
\begin{equation}
\det(\eta_{ij})_{u}=T^{4}\det(\eta_{ij})_{s}
\end{equation}
\par
It is important to emphasize relation $(4.23)$ because of the
factor $T^{4}$ which takes us to a possible connection with the
Stefan-Boltzmann Law.
\par
\section{Conclusions}
\par
With this paper we have tried to make a possible connection
between thermodynamic length and work using both energy and
entropy metric. The possible interpretation that we have found is
that for Ideal and quasi-Ideal gas, thermodynamic length
"measures" the amount of work, along isotherms, that a
thermodynamic system does to keep its ideal state or its
equilibrium.
\par
\section{Acknowledgments}
\par
I gratefully thank Prof. P. Salamon for reading my manuscript and
for replying with very useful comments. I thank Prof. R.S. Berry
for his patience in replying to my e-mails and I also thank Prof.
S. Preston who gave me many useful advices.
\par

\end{document}